\documentclass[fleqn,10pt]{wlscirep}
\usepackage[utf8]{inputenc}
\usepackage[T1]{fontenc}
\usepackage{hyperref}
\usepackage{float}
\usepackage{amsmath}

\DeclareMathOperator*{\argmin}{arg\,min}
\title{Freecyto: Quantized Flow Cytometry Analysis for the Web}
\author[1,*]{Nathan Wong}
\author[1]{Daehwan Kim}
\author[1]{Zachery Robinson}
\author[1]{Connie Huang}
\author[1,*]{Irina M. Conboy}

\affil[1]{Department of Bioengineering and QB3, UC Berkeley, Berkeley, CA 94720, USA}

\affil[*]{nathanwong@berkeley.edu}
\affil[*]{iconboy@berkeley.edu}

\begin{abstract}
Flow cytometry (FCM) is an analytic technique that is capable of detecting and recording the emission of fluorescence and light scattering of cells or particles (that are collectively called “events”) in a population\cite{ONeill2013}. A typical FCM experiment can produce a large array of data making the analysis computationally intensive\cite{Lugli2010}.  Current FCM data analysis platforms (FlowJo\cite{FlowJo}, etc.), while very useful, do not allow interactive data processing online due to the data size limitations.  Here we report a more effective way to analyze FCM data. Freecyto is a free, easy-to-learn, Python-flask-based web application that uses a weighted k-means clustering algorithm to facilitate the interactive analysis of flow cytometry data. A key limitation of web browsers is their inability to interactively display large amounts of data. Freecyto addresses  this bottleneck through the use of the k-means algorithm to quantize the data, allowing the user to access a representative set of data points for interactive visualization of complex datasets. Moreover, Freecyto enables the interactive analyses of large complex datasets while preserving the standard FCM visualization features, such as the generation of scatterplots (dotplots), histograms, heatmaps, boxplots, as well as a SQL-based sub-population gating feature\cite{Lugli2010}. We also show that Freecyto can be applied to the analysis of various experimental setups that frequently require the use of FCM. Finally, we demonstrate that the data accuracy is preserved when Freecyto is compared to conventional FCM software. 
\end{abstract}
\begin{document}

\flushbottom
\maketitle
%
%
\thispagestyle{empty}
\section*{Keywords}
Flow cytometry, Big data analysis, Web application, Machine learning, Unsupervised learning, Data Quantization, Software development

\section*{Introduction}
\begin{flushleft}
Flow cytometry is broadly used in biomedicine, which is exemplified by identification of protein marker expressions\cite{RAMEL19921220,Leith,Rosner2013}, determinations of cell-fate and cell cycle progression\cite{Darzynkiewicz1992}, analysis of pathology-caused changes, e.g. cancer promoted, immune-skewing, etc. \cite{Barlogie3982,Keyes2020,Brando,Lugli2009}, testing therapeutic efficacy of a treatment\cite{Benedek2014}, and, more recently, gene-editing detection workflows\cite{Hu2014}. A common experimental setup in biomedicine relies on being able to identify specific changes between a control and an experimental cell population. The changes between control and experimental cohorts are often determined through fluorescently tagged antibodies that are specific for given proteins; and the fluorescence is examined by microscopy and/or high throughput screening using a flow cytometer\cite{ONeill2013,McKinnon2018}. 
\end{flushleft}

\begin{flushleft}
Successful FCM experiments rely on the accuracy and resolution of the data analysis, e.g. the performance of the FCM software that provides quantitative outputs for large numbers of events\cite{Lugli2010}. In FCM analysis, an event is constituted by the cytometer’s detection of fluorescence emission and/or light scatter signals from a single cell or particle that passes through the microfluidic flow chamber. With thousands of these events, individual measures of fluorescence, size and granularity are produced, and to add complexity, these measurements can be deliberately modified by a researcher through the instrument setup, which can be changed from run to run\cite{Maecker2006}. FCM analysis, thus, becomes a computational and statistical challenge that produces meaningful data only if the analysis is adequate for the experimental complexity. Inherent in this requirement, the datasets that are produced with the conventional FCM software (FlowJo\cite{FlowJo}, Cytobank\cite{Kotecha2010}, OpenCyto\cite{Finak2014}, and Webflow\cite{Hammer2009}) are typically quite large, which complicates their interactive web analyses. 
\end{flushleft}
\begin{flushleft}
In this work we developed a new FCM software that facilitates the FCM data analysis, while maintaining the accuracy and resolution of the data. In fact, analysis of flow cytometry experiments, despite having tens of thousands of data points, can be performed and visualized on a mobile device. Importantly, while simplifying the data analysis and having the intuitive work flow,  Freecyto preserves the key features of traditional FCM software, such as scatterplots (dotplots) of two different emission, histograms of a fluorescent emission measurement\cite{McKinnon2018}, the side-by-side comparison of the results between the control and experimental populations and gating on sub-populations of cells. 
\end{flushleft}

\begin{flushleft}
Similarly to FlowJo\cite{FlowJo}, Cytobank\cite{Kotecha2010}, OpenCyto\cite{Finak2014}, and Webflow\cite{Hammer2009}, Freecyto supports machine learning applications, but it does not require the installation of specific software packages (often OS-dependent), a detailed understanding of the software workflow, or extra layers of complexity in displaying, interacting, and sharing the FCM analysis with other researchers. Additional features of Freecyto are robust data-management and data-sharing: Freecyto is built on a secure centralized database management system, allowing for data to be stored remotely and analyses to be shared and edited by anyone, yet it maintains the safeguard of proper permissions.  Notably, the decisions on instrument settings (such as, changing the gain and signal intensity) and experimental set-ups (for instance, additional runs of certain cohorts) become better informed - based on real time user-friendly data analysis.
\end{flushleft}
\begin{flushleft}
A key feature of Freecyto is the k-means clustering algorithm in which data points are clustered together into k clusters based on a Euclidian distance metric. This use of k-means algorithm as a method of data quantization is distinct from the flow cytometry studies, which use clustering algorithms to analyze the data\cite{Murphy1985,Bruggner2014,Ye2019,Ge2012,Dorfman2016}. Freecyto, in contrast, uses k-means to create a reduced, representative dataset of the original, so that the user can have much greater capability in analyzing the data, such as applying the stated clustering algorithms to the data. The original data is then reduced to the centers of the clusters, allowing the user to gate interactively on these centers. We show that FCM data analysis remains faithful when Freecyto is compared to the conventional FlowJo software. 
\end{flushleft}
\begin{flushleft}
By focusing and quantizing the data, Freecyto offers a better control over the analysis of FCM experiments, increasing the computational feasibility of any and particularly, very large datasets. Because of the high dimensional nature of flow cytometry data and the increasing technological developments in flow cytometers which have pushed the number of parameters and the sheer volume of data ever higher, there is a greater need for FCM software to handle increasingly large data sets\cite{Bendall2011,Mair2015}. Freecyto was developed to address this challenge.
\end{flushleft}

\section*{Results}
\subsection*{K-means Quantization}
\begin{figure}[!htbp]
\centering
\includegraphics[width=0.9\linewidth]{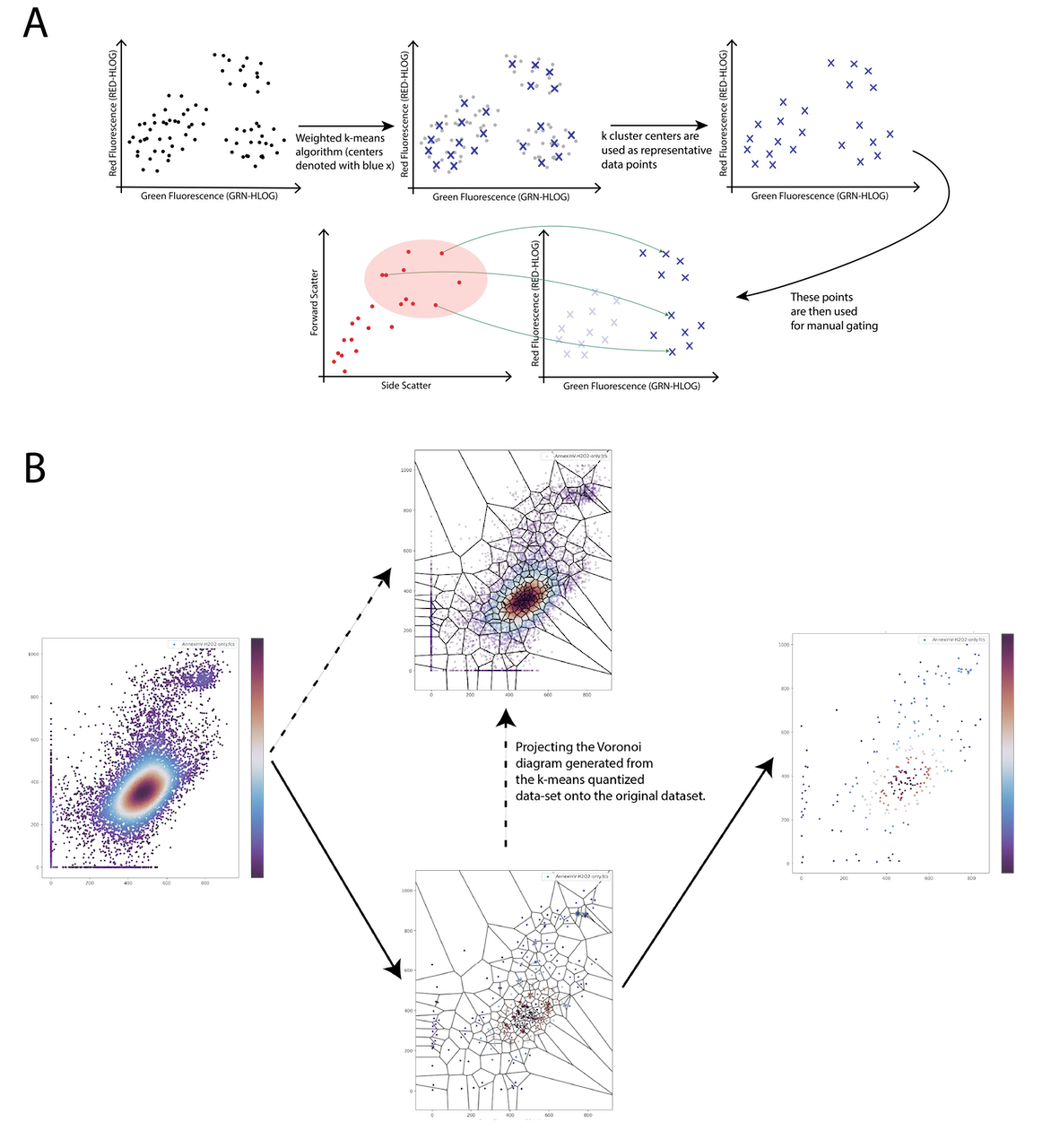}
\caption{\textbf{K-means Workflow in Freecyto.} \textbf{(A)} The process by which the original dataset is quantized, and how manual gating works on a shared data source. \textbf{(B)} The principles behind k-means quantization, and the Voronoi diagram computed from the reduced dataset projected on the original dataset.}
\label{fig:f1}
\end{figure}
\begin{flushleft}
While the quick visualization capabilities are sufficient for most basic flow cytometry operations, a more detailed study may require additional specialized functions, such as sub-population gating and quadrant (coordinate-system) gating. Having data sets on the magnitude of $10^5$ or $10^6$ events, presents a significant challenge to interactively plot these on the web. In the case of gating, having tens of thousands of points that users can lasso-select on the web is virtually impossible for personal computers and standard web browsers. Freecyto solves this problem by introducing a k-means clustering algorithm for quantizing the input data (Figure \ref{fig:f1}).
\end{flushleft}

\begin{flushleft}
First, after running the k-means clustering algorithm, the centroids are used to construct a Voronoi diagram. Thus, the original dataset is partitioned into Voronoi cells, and each cell contains all the original points that belong to that cluster. Following, for each Voronoi cell, the variance is computed, with the centroid used as the mean of the geometric space. Finally, the within-cluster variance is plotted as a colormap within the Voronoi diagram to portray which cells contain more of the underlying variance, and the variance is summed up across all Voronoi cells to portray the elbow at which minimal within-cluster variance is lost with respect to the increase in computation power due to increasing the number of clusters.
\end{flushleft}

\begin{flushleft}
K-means clustering (implemented with Lloyd’s algorithm, clusters initialized with kmeans++ with a default seed) is an unsupervised machine-learning algorithm that is used to identify clusters of points based on each point’s distance from the center of a proposed cluster. Freecyto runs this algorithm on the user-selected channels, identifying a pre-defined number of clusters, and storing only the centers of these clusters. The number of clusters is either user-selected (if running locally) or approximated automatically as a range between 250 and 5000 based on the size of the dataset. This simplifies the conventional k-clustering approach and enables future development of more suitable algorithms to determine k\cite{Yuan2019,Pham2005}. Freecyto’s application of k-means clustering quantization vastly reduces the complexity of the flow cytometry data, without significant loss to the variability within the original dataset as we will show in the next section. The reduced dataset that is generated is highly suitable for downstream statistical analysis, such as hierarchical clustering or dimensionality reduction to identify sub-populations of cells (Supplemental Figure 5). 
\end{flushleft}

\subsection*{Fidelity of Data Quantization in Interactive Analysis.}
\begin{figure}[!htbp]
\centering
\includegraphics[width=0.8\linewidth]{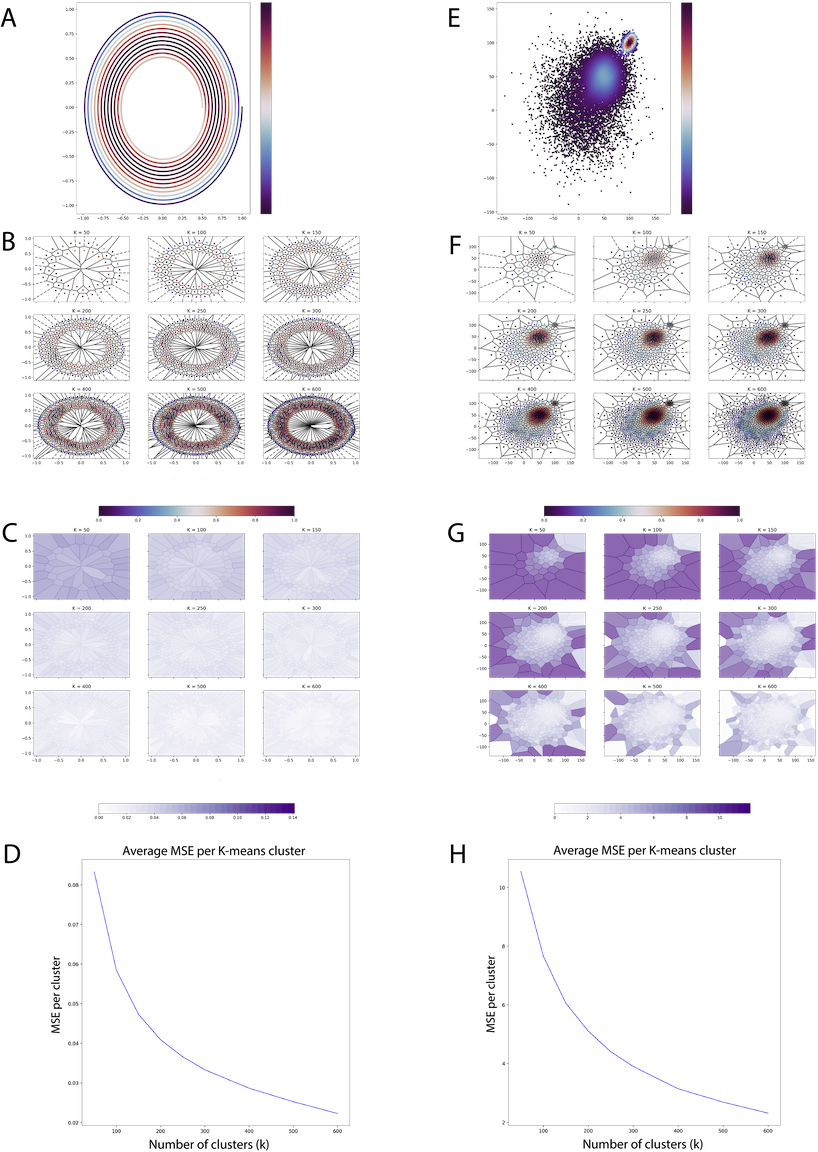}
\caption{\textbf{K-means Within-Cluster Variance Visualization of Synthetic Datasets.} \textbf{(A)} Original spiral data (N=5000). \textbf{(B)} Cluster centers with Voronoi cells outlined. \textbf{(C)} Within-cluster variance of each Voronoi cell with increasing k, and by extension, the MSE in each cluster identified by k-means. \textbf{(D)} Trend of increasing clusters and the average within-cluster variance of each cluster. \textbf{(E)} Original bimodal data (N=10000). \textbf{(F, G, H)} Cluster centers and variance loss in each Voronoi cell with increasing k.}
\label{fig:f2}
\end{figure}
\begin{flushleft}
To quantitively examine the quality of our reduced data set, we compute the mean-squared error (MSE) of each cluster. For the k-means algorithm, this is equivalent to computing the within-cluster variance of each cluster, because the predicted cluster center is the mean of all points in that cluster. The MSE of each cluster, as visualized by Voronoi cells, is then mapped to a color range to depict how faithfully each cluster center captures the other points in that cluster. In Figure 2C and 2G, it’s shown that with increasing k, the lower the MSE for each cluster. Finally, the average of all the MSE for all clusters is computed (2D and 2H) to show that the data lost in each cluster center decreases rapidly in exchange for smaller increases in the number of clusters chosen. 
\end{flushleft}
\begin{flushleft}
The quantized data can then be plotted interactively through Bokeh on a webpage and downloaded as a SQL database within the web application. In this interactive analysis portion, each flow cytometry data file is treated as a shared data source, thus in Freecyto the user can lasso-select a sub-population of cells that are displayed in a scatterplot graph or a fluorescence channel and observe the quantized data for that sub-population of cells in the other FCM channel(s). This Freecyto feature allows the user to quickly and with more precision determine how the size of the cells or a signal for a specific marker (cell-fate protein, for example) is related to other markers (transgene expression, for instance) for each cell in the studied population. Demo: (\href{https://www.youtube.com/watch?v=JlIVgxh4_YA&feature=youtu.be&t=187}{3:07} – \href{https://youtu.be/JlIVgxh4_YA?t=380}{6:20})
\end{flushleft}

\begin{figure}[!htbp]
\centering
\includegraphics[width=0.9\linewidth]{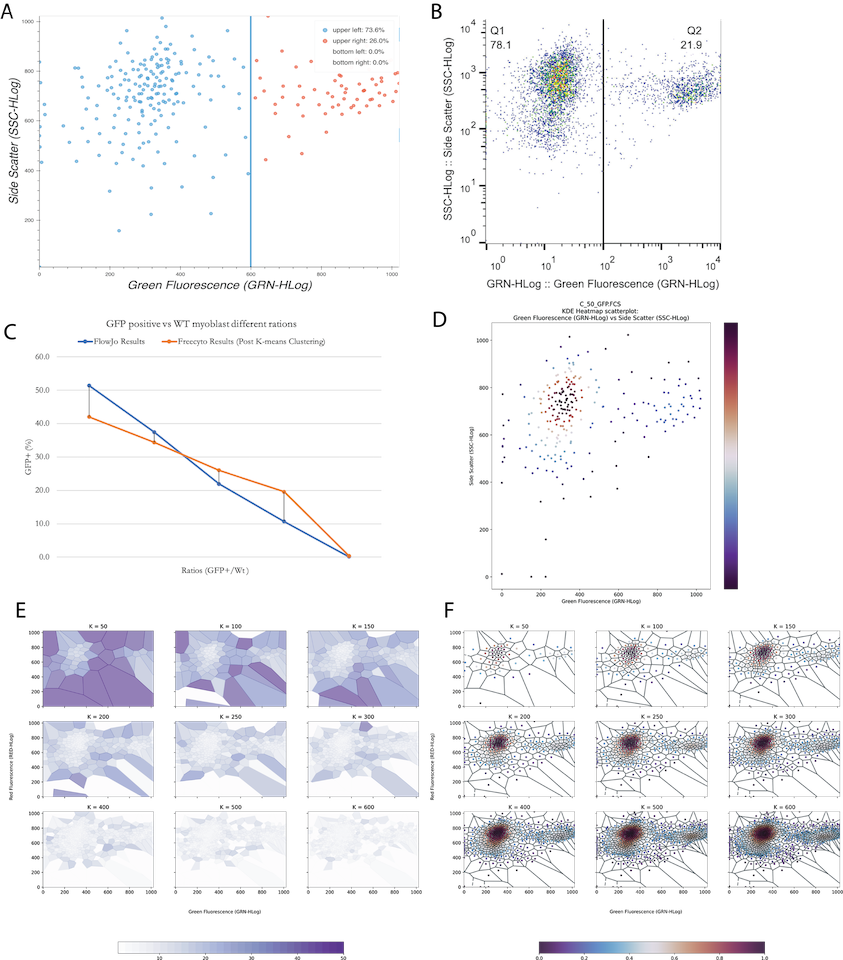}
\caption{\textbf{Analysis of GFP positive and negative cell populations.} \textbf{(A)} 50:50 GFP transgenic cells ratios with the coordinates gated by Freecyto (after quantization).\textbf{(B)} The same 50:50 GFP transgenic cell ratios with the coordinates gated by FlowJo. \textbf{(C)} Compares Freecyto and FlowJo measurements of GFP+ cells for 100:0, 75:25, 50:50, 25:75, and 0:100 ratios. \textbf{(D)} Density plot created by Freecyto which outlines the density of cells after the k-means quantization is performed with 250 clusters. \textbf{(E)} MSE of each cluster with varying values of k. \textbf{(F)} The resulting density plot with varying values of k.}
\label{fig:f3}
\end{figure}

\begin{figure}[!htbp]
\centering
\includegraphics[width=0.8\linewidth]{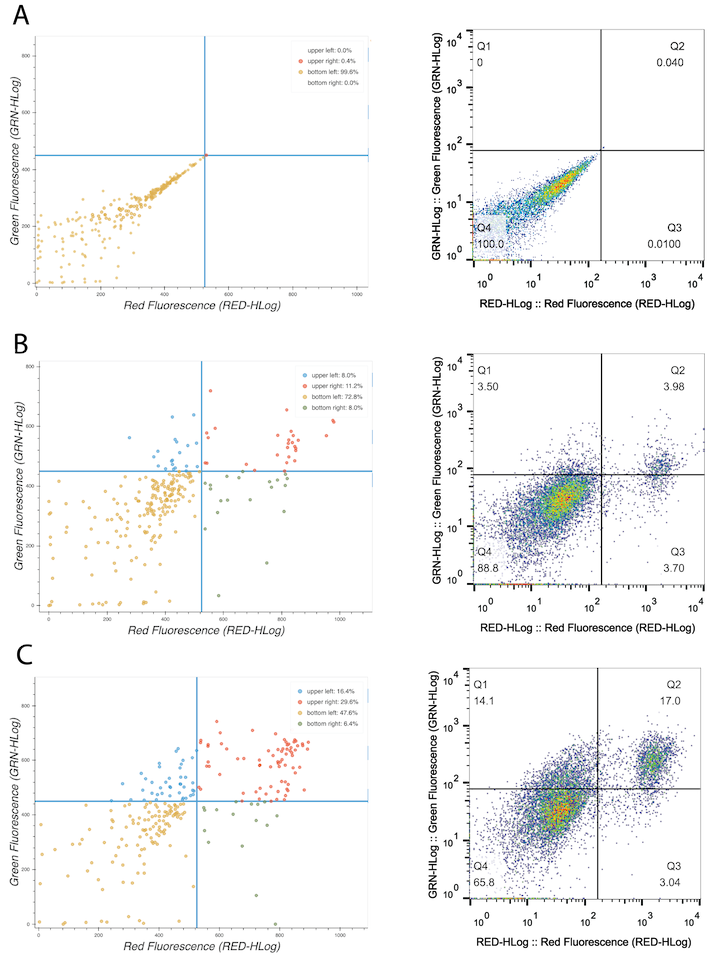}
\caption{\textbf{Analysis of Apoptosis.} IMR90 cells were treated with hydrogen peroxide, H$_2$O$_2$, at 200 $\mu$M for 24h to induce apoptosis. The cells were then stained with Annexin V and 7-AAD. Early apoptotic cells are positive for Annexin V and are seen in the top left quadrant (Q1) and late apoptotic cells, which are positive for both annexin and 7-AAD are seen in the top right quadrant (Q2). Live cells are negative for both stains (Q4). \textbf{(A)} Negative control: Isotype-matched IgG staining (1st antibody) + secondary (FITC). \textbf{(B)} Untreated group. \textbf{(C)} H$_2$O$_2$ treatment group.}
\label{fig:f4}
\end{figure}

\begin{flushleft}
One key question is whether our method of k-means clustering qualitatively maintains the accuracy and resolution of the data. To address this, we compared side-by-side Freecyto and the conventional FCM software FlowJo in the analysis of GFP positive cells in a population and in studying cells in early and late stages of apoptosis (e.g. AnnexinV-7AAD and co-stain).  Here we used Freecyto modality for such a common feature of FCM as a coordinate system gating to identify the percentage of cells located within certain thresholds. As shown in Figures \ref{fig:f3} and \ref{fig:f4}, Freecyto was as accurate as FlowJo in the resolution of these data sets, at the same time preserving the key features of FCM software, such as allowing the user to specify fluorescence thresholds and visualize and quantify the percentage of cells located in these quadrants (Figures \ref{fig:f3}, \ref{fig:f4}).
\end{flushleft}
\begin{flushleft}
Moreover, Freecyto generated quantized data points are stored in an SQLite database - essential to the deep gating tool. The deep gating tool allows the user to lasso-select a sub-population of cells and graphically display only the gated cells for all advanced analysis operations. This is useful in narrowing the analysis to specific sub-populations, as well as identifying outliers in the dataset. This deep-gating function can be applied as many times as needed, and all deep-gates can be reset by pressing the reset-gating button, after which the visualization and quantification of the results will reflect the original, unaltered dataset (Figures 3, 4). Both the results of the k-means quantization and the sub-populations identified from manual gating can be downloaded directly in the application.
\end{flushleft}
\begin{flushleft}
To comparatively analyze the accuracy and capabilities of Freecyto and FlowJo, WT and GFP+ cells were mixed at five different ratios, 100:0, 75:25, 50:50, 25:75, and 0:100, WT:GFP+; and run on Guava Easycyte Flow cytometer (Millipore-Sigma). The data was analyzed by FlowJo and Freecyto in parallel. As a result, the number of GFP positive cells increased linearly from 100:0 WT/GFP+ to 0:100 WT/GFP+, as expected, which was accurately detected by both FlowJo and Freecyto. 
\end{flushleft}
\begin{flushleft}
To compare Freecyto and Flowjo in another commonly analyzed by Flow Cytometry assay – cell apoptosis, IMR90 human fibroblasts were treated (or not) with hydrogen peroxide, H$_2$O$_2$, at 200 $\mu$M for 24h to induce apoptosis.  The cells were assayed with Annexin V and 7-AAD and run on the Guava Easycyte Flow cytometer (Millipore-Sigma).  The results were analysed with Freecyto, yielding accurate and visually clear data. The negative control, isotype-matched IgG fluorescence was used to set up the quadrant, Figure 4A. Early apoptotic cells positive for Annexin V can be seen in the top left quadrant and late apoptotic cells positive for both Annexin V and 7-AAD in the top right quadrant. As expected, Freecyto shows the number of Annexin V positive cells, Figure 4B. The number of cells in early and late stages of apoptosis were increased with H$_2$O$_2$, as compared to the untreated control, Figure 4C. In summary, the analysis of apoptosis (Annexin V and 7ADD assay) yields the predicted results and is as accurate and sensitive with Freecyto as it is with Flowjo.
\end{flushleft}

\subsection*{Web (Uwsgi-flask-nginx) application to allow platform-agnostic, mobile-ready access to flow cytometry analysis}
\begin{figure}[h!]
\centering
\includegraphics[width=1\linewidth]{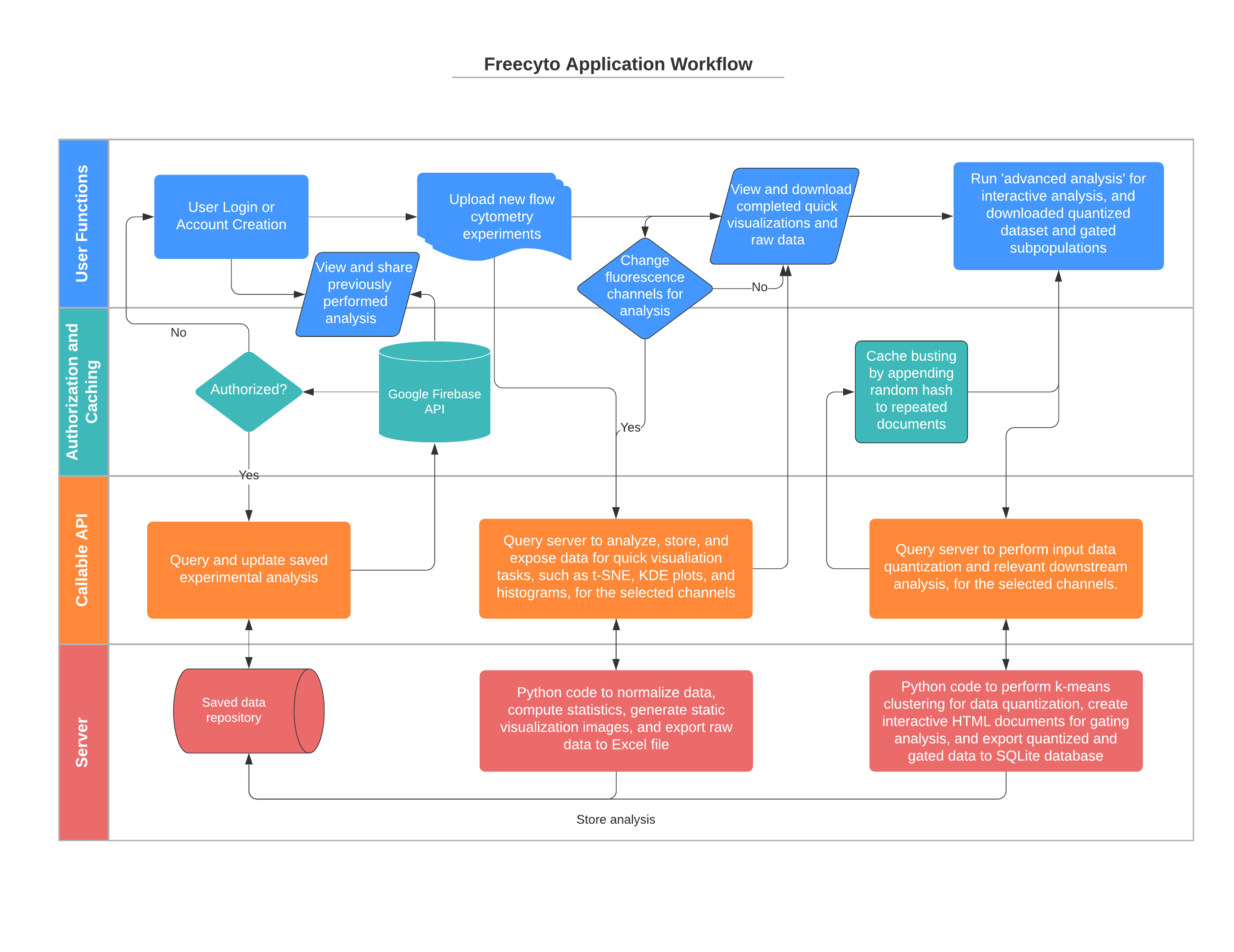}
\caption{Freecyto Application Workflow}
\label{fig:f5}
\end{figure}
\begin{flushleft}
Several core technologies are deeply integrated into Freecyto in order to allow seamless processing and visualization of flow cytometry data. Chiefly, the integration of these technologies allows for robust storage of user data, high-throughput handling of the data, e.g. processing operations, and interactivity of the data visualizations. 
\end{flushleft}
\begin{flushleft}
Computationally expensive operations in flow cytometry, including reading and parsing data, performing visualizations, and obtaining sample statistics, are all performed server-side in Freecyto. Freecyto is hosted as a Python-flask-uwsgi-nginx application on a Digital Ocean server.
\end{flushleft}

\begin{flushleft}
While most flow cytometry tools have unique requirements depending on the user’s operating system (OS), application dependencies (a specific version of python packages), or computational resources (i.e. four CPU cores), Freecyto can be accessed without platform restrictions and dependencies. This application also is designed to be mobile-compatible, allowing users to access their flow cytometry analysis and also perform new flow cytometry analysis directly on their mobile devices (Figure \ref{fig:f5}).
\end{flushleft}
\begin{flushleft}
In addition, Freecyto can be downloaded as a Flask application (open-source), so that users can install the appropriate dependencies and run the application on a local intranet (useful if users desire a stricter control of Flow cytometry data privacy). This also allows for greater control over default parameters and application modules, such changing the number of reduced data points used in interactive analysis and implementing a clustering model on top of the reduced data set (Figure 5). 
\end{flushleft}
\begin{flushleft}
Demo: (\href{https://youtu.be/JlIVgxh4_YA?t=0}{0:00} – \href{https://youtu.be/JlIVgxh4_YA?t=60}{1:00})
\end{flushleft}

\subsection*{Parallel processing (multiprocessing) of computationally intensive analysis functions}

\begin{flushleft}
Freecyto integrates advances in multiprocessing functionality in order to speed up traditionally expensive FCM data analysis operations. Multiprocessing is implemented when users upload multiple files, when visualizations are performed, and when the k-means algorithm is running. These operations are asynchronously performed on the server-side, speeding up the time it takes for the user to receive analyses outputs from their data by an order of magnitude. Through the implementation of this multiprocessing a side-by-side over five files upload becomes possible (Supplemental Figure 3).
\end{flushleft}
\subsection*{User data management and authentication}

\begin{flushleft}
Google Firestore/Datastore is integrated to store references to previously performed visualization operations. For example, the images that are generated from an experimental upload are stored in a unique directory on the server, and the references to the generated images are stored in a collection as a unique entry under the user account in Google Firestore. This prevents redundant analysis operations (i.e. the user uploads the same experimental files), yet, it allows the user to access the previously performed operation. A sortable table of previously performed experiments (50 most recent) are listed in the user home page, allowing the user to easily access previously analysed flow cytometry results. 
\end{flushleft}

\begin{flushleft}
Firebase and Google identity platform: Google and Email logins are enabled, allowing the user to create and access their user account with these authentication methods. This prevents unauthorized usage of the application, requiring the user to create an account before accessing the analysis toolkit. To promote scientific knowledge and collaborations, sharing the results of a flow cytometry experiment on Freecyto merely requires sharing the URL of the experiment. Demo: (\href{https://youtu.be/JlIVgxh4_YA?t=60}{1:00} – \href{https://youtu.be/JlIVgxh4_YA?t=90}{1:30})
\end{flushleft}
\subsection*{Side-by-side experiment comparisons (multiple file upload)}
\begin{flushleft}
Freecyto supports user upload of multiple flow cytometry files as a result of the multiprocessing pipeline. For normalization of the raw input files, the user may select hyperlog, logicle, or no transformation to be applied. Logicle and hyperlog transformations normalize the flow cytometry data by transforming most events (including negatively measured values) to a normalized fluorescence value of between 0 and 1\cite{Bagwell2005}. This improves on traditional free flow cytometry analysis applications, which limit the user to uploading only a single flow cytometry file at a time, though many flow cytometry experiments have anywhere from 2 to 10+ files to analyse. Freecyto’s approach allows the user to upload numerous files concurrently, enabling plots to be overlaid for easy and clearly visualized comparison between the datasets. In another feature of Freecyto, if overlays make it harder to discern the individual plots, then individual files can also be graphed and visualized.  Demo: (\href{https://www.youtube.com/watch?v=JlIVgxh4_YA&feature=youtu.be&t=90}{1:30} – \href{https://www.youtube.com/watch?v=JlIVgxh4_YA&feature=youtu.be&t=120}{2:00})
\end{flushleft}
\subsection*{Quick visualization capabilities}
\begin{flushleft}
Freecyto is built on the principle that FCM analysis should be easy to perform and that real-time data processing expands the research capabilities in acutely and accurately modulating the FCM experiments. Freecyto’s pipeline achieves this by quick visualization of the scatterplots, density-estimation plots, histograms, box-whisker diagrams, and correlation tables, which are generated by Freecyto based on the selected fluorescence channels. In addition, t-SNE plots allow users to visualize segregating features of the data. The images and relevant statistics are displayed through a carousel slider (Siema) and a table respectively.
\end{flushleft}
\begin{flushleft}
It is integral to flow cytometry analysis to allow users to select the fluorescence channels they wish to visualize. Freecyto accomplishes this with a simple checkbox list of all possible channels. The user selects the channels they wish to visualize, presses “submit,” and the images automatically update to match the desired fluorescence channels to visualize. This pipeline is designed to be minimalistic – it allows the user to quickly determine how their data looks, offering enough modularity to facilitate the most common flow cytometry analysis operations. In addition, the converted flow cytometry data can be downloaded as an Excel spreadsheet. Demo: (\href{https://www.youtube.com/watch?v=JlIVgxh4_YA&feature=youtu.be&t=120}{2:00} – \href{https://www.youtube.com/watch?v=JlIVgxh4_YA&feature=youtu.be&t=187}{3:07})
\end{flushleft}

\section*{Discussion}

\begin{flushleft}
Freecyto was developed as a new data processing software for Flow Cytometry data and validated for enhancing the speed, convenience, and machine learning capacity of the FCM data analysis, while preserving the accuracy. These features were validated in key FCM set-ups of studying sub-populations with variable expression of a transgene, and in viability-apoptosis studies. Summarily, the use of our weighted k-means clustering algorithm innovated FCM data analysis and transformed it into a simple, easy to use online platform. 
\end{flushleft}
\begin{flushleft}
Freecyto offers all the necessary features to perform typical FCM analyses, in addition to providing the user interactive analysis of the data and it fills a niche when compared with other FCM software (Table 1). Freecyto is a flexible platform that allows modifications. For example, Opencyto allows users to create automated gating pipelines in R which may solve the subjectivity and time-consuming nature of manual gating and such a feature is very compatible to build on top of Freecyto’s existing framework\cite{Finak2014}. Freecyto does not innovate the existing flow cytometry analysis, instead it innovates the approach to such analyses, thereby improving on the ease and accessibility of FCM data, while also providing greater flexibility and control in gating large datasets, through the quantizing of the data with a weighted k-means clustering algorithm.
\end{flushleft}
\begin{table}[h!]
\centering
\begin{tabular}{|p{0.1\linewidth}||p{0.2\linewidth}|p{0.2\linewidth}|p{0.2\linewidth}|p{0.2\linewidth}|}
\hline
\textbf{Feature} & \textbf{Freecyto} & \textbf{Opencyto} & \textbf{Cytobank} & \textbf{FlowJo} \\
\hline
\hline
\textbf{What is it?} & Python web application & R software package & Cloud-based web server & Software package (OS dependent)\\
\hline
\textbf{Summary} & K-means algorithm allows interactive gating between any combination of channels in side-by-side graphs & Pipeline for automated gating algorithms (as opposed to manual gating) & Specialized service that uses many different tools e.g Citrus\cite{Bruggner2014} to perform FCM analyses & Automation of repeated analyses, customizable data visualizations\\
\hline
\textbf{Free-to-use} & Yes & Yes & No & No \\
\hline
\textbf{Requires software download} & No & Yes & No & Yes \\
\hline
\textbf{Straight-forward data analysis sharing} & Yes & No & Yes & No \\
\hline
\textbf{Beginner-friendly} & Yes & No & No & No \\
\hline
\textbf{Mobile-compatible} & Yes & No & No & No \\
\hline
\end{tabular}
\caption{\label{tab:example}Comparing Freecyto with other flow cytometry applications.}
\end{table}

\section*{Conclusions}
\begin{flushleft}
FCM analysis is essential for a broad range of biomedical studies, many of which are directly and critically important for human health.  Freecyto allows for the streamlined, fast, facile, user-friendly and easy to share analysis of multiple FCM experiments in parallel, harnessing the transmissibility of internet ease-of-use to power and serve its analytical platform. Whereas many FCM analysis packages are expensive, require software/OS dependencies, or have a significant learning curve, Freecyto is free, web-based, and easy to use, and while simplifying FCM studies, Freecyto improves the processing of high-volume data and facilitates the real-time data analysis. 
\end{flushleft}
\begin{flushleft}
As flow cytometry development continues to improve, the need for indexing and manipulating large quantities of scientific data cannot be understated. Freecyto integrates state-of-the-art data storing and indexing features with Google Cloud, creating an interface for users to have greater confidence and connectivity with their flow cytometry data. In this regard, our k-means quantization approach might be broadly useful and important not only in FCM, but more broadly, for Big Data analysis in omics, medical data for machine learning and AI, computer vision, environmental engineering, etc. large data realms.
\end{flushleft}

\section*{Materials and Methods}
\subsection*{Data Visualization}
Several Python packages were used in creating this application. Flask was used to serve the web application. Google Identity (Firebase) was used to authenticate users, and Google DataStore was used to store references to previously performed experiments. Pandas, NumPy, FlowUtils, and Cytoflow were used to dynamically store and transform the raw flow cytometry data. Matplotlib, Seaborn, and Pandas were used to generate images of scatterplots, box-plots, heatmaps, and histograms. The t-distributed stochastic neighbour embedding (t-SNE) projection was performed with Scikit-learn (sklearn) with perplexity of 40. For the interactive analysis, sklearn was used for the weighted k-means clustering. SQLite3 was used to store clustered data. Bokeh and Holoviews were used to display the interactive graphs. HTML5UP and Creative Tim Light Bootstrap Theme inspired the front-end template design of the web application.
\subsection*{Multiprocessing}
Multiprocessing, assuming a multi-core machine, was implemented to speed up the data visualization algorithms. Chiefly, the results of a benchmark test on a quad-core, 8 GB RAM, 2.3 Ghz MacBook Pro are reported below for the static image visualizations, and for the interactive data analysis portions.
\subsection*{Weighted K-means Algorithm}
$X= \{x_1,x_2,...,x_n\}$ such that every $x_i$ has $d$ dimensions. Let $\Omega$ be a diagonal $d$ x $d$ matrix such that the diagonal entries are the weights of each dimension. $k$ is the number of clusters we want to find. $S$ is the set of all $k$ clusters such that $S=\{S_1, S_2,...,S_k
\}$. We want to minimize the loss function:
$$\argmin_S \sum^k_{i=1} \sum_{x \in S_i} (x-\mu_i)^T\Omega (x-\mu_i)$$

\noindent In the default case, let the diagonal entries of $\Omega$ be $1$ if the corresponding channel was selected for visualization, and $0$ otherwise. 

\subsection*{Voronoi Diagram Algorithm}
$X= \{x_1,x_2,...,x_n\}$ such that every $x_i$ has $d$ dimensions. $R$ is the set of all $k$ Voronoi diagrams such that $R=\{R_1, R_2,...,R_k
\}$ and $S$ is the set of all $k$ clusters such that $S=\{S_1, S_2,...,S_k
\}$. $d$ is a distance metric, for which we used Euclidean distance. We want to find the region such that every point in the region is closest to the set of points described by the k-means clustering. 
$$R_k = \{x \in X | d(x, S_k) \leq d(x, S_j) \forall j \neq k\}$$
Or equivalently, because the distance of every point x in $S_k$ to it’s mean centroid $\mu_k$ has already been minimized in the converged k-means algorithm: 
$$\forall x \in S_k | d(x, S_k) \leq d(x, S_j)$$
$$\forall j \neq k \implies R_k = \{x \in S_k\}$$

\subsection*{Web application (open-source) licenses}
\begin{itemize}
\item Advanced Analysis: Light bootstrap theme by Creative Tim: MIT License 
\\https://github.com/timcreative/freebies/blob/master/LICENSE.md
\item Lens by HTML5UP: Creative Commons 3.0 https://html5up.net/license
\item NumPy: https://github.com/numpy/numpy/blob/master/LICENSE.txt 
\item SciPy: https://scipy.org/scipylib/license.html 
\item Scikit-learn: https://scikit-learn.org/stable/
\item Pandas: https://github.com/pandas-dev/pandas/blob/master/LICENSE 
\item Matplotlib: https://matplotlib.org/users/license.html 
\item Bokeh: https://github.com/bokeh/bokeh/blob/master/LICENSE.txt 
\item Holoviews: https://github.com/pyviz/holoviews/blob/master/LICENSE.txt
\item Flask: http://flask.pocoo.org/docs/1.0/license/ 
\item SQLAlchemy: https://docs.sqlalchemy.org/en/latest/copyright.html 
\item Cytoflow: https://github.com/bpteague/cytoflow/blob/master/LICENSE.txt 
\item FlowUtils: https://github.com/whitews/FlowUtils/blob/master/LICENSE
\end{itemize}

\subsection*{Myoblast cultures}
Transgenic GFP+ and WT (C57.B6) mouse myoblasts were cultured in growth medium: Ham’s F10, 20\% Bovine Growth Serum and 5 ng/ml bFGF on 1 $\mu$g/cm$^2$ Matrigel.  Cells were washed and detached with PBS (three 37C) and were pelleted by centrifugation. Cells were pelleted and counted using a hemocytometer. 

\subsection*{Cell culture and apoptotic assay}
Normal human lung fibroblast cells (IMR-90) were obtained from ATCC \#CCL-186. Cells were maintained in DMEM (Dulbecco’s Modified Eagle Medium) supplemented with 10\% fetal calf serum (FCS, Hyclone) containing 1\% penicillin-streptomycin (Invitrogen) and maintained in a humid atmosphere at 37°C containing 5\% CO$_2$. When cells were grown to 70\% confluence, they were subcultured at $\frac{1}{5}$ dilution for later passaging.

\begin{flushleft}
 The apoptotic assay of IMR90 was conducted by Apoptosis Detection Kit (ab214663, Abcam) according to the manufacturer’s protocol. Briefly, cells were detached using 0.05\% trypsin and washed twice with PBS. Then, samples were resuspended in 1x annexin-binding buffer and incubated with 5 $\mu$L Annexin V-FITC and 5 $\mu$L 7-amino-actinomycin D (7-AAD) for 15 min at 37°C, avoiding light. Finally, events were acquired with a Guava Easycyte Flow cytometer (Millipore-Sigma) and analysed by Freecyto and Flowjo software individually to quantify the distribution of cells. 
\end{flushleft}

\section*{Abbreviations} 
\textbf{FCM}: Flow cytometry \\
\textbf{Event(s)}: Emission(s) of fluorescence and light scattering of cells or particles\\
\textbf{t-SNE}: Barnes-Hut approximation of t-distributed stochastic neighbour embedding\\
\textbf{K-means}: Lloyd’s Algorithm with Euclidean distances for k-means clustering (k-means++ is used for cluster center initialization).\\
\textbf{MSE}: Mean squared error\\
\textbf{WT}: Wild type \\
\textbf{GFP}: Green fluorescent protein \\
\textbf{IMR-90}: Human lung fibroblast cells

\section*{Data Availability} 
\begin{flushleft} 
The datasets generated and/or analysed during the current study are available in the Freecyto Github repository, \href{https://github.com/nathan2wong/freecyto/tree/master/datasets}{https://github.com/nathan2wong/freecyto/tree/master/datasets}. 
\end{flushleft}
\noindent Project name: Freecyto\\
Project homepage: \href{https://freecyto.com}{https://freecyto.com}\\
Demo: \href{https://youtu.be/JlIVgxh4_YA}{https://youtu.be/JlIVgxh4{\_}YA} \\
Archived version: \href{https://github.com/nathan2wong/freecyto}{https://github.com/nathan2wong/freecyto} \\
Operating system(s): Platform independent\\
Programming Language: Python, JavaScript\\
Other requirements: Listed on GitHub\\
License: BSD3\\
Any restrictions to use by non-academics: License Needed	

\section*{Acknowledgements}

We would like to thank Alex Park for providing technical help with these studies, and Michael Conboy for the helpful suggestions on the work and the manuscript. 
\section*{Funding}
This work was supported by NIH R01 EB023776, R01 HL139605 and Open Philanthropy awards to IC, and the funds were used to support the data collection of the study.

\section*{Author Information}

\subsection*{Affiliations}
Department of Bioengineering and QB3, UC Berkeley, Berkeley, CA 94720, USA \newline
Nathan Wong, Daehwan Kim, Zachery Robinson, Connie Huang, and Irina M. Conboy

\subsection*{Contributions}
NW created the Freecyto software and wrote the manuscript. ZR provided figures, data, and analyses of the GFP cell experiment (Figure 3). DH provided figures, data, and analyses of the apoptotic cell experiment (Figure 4). CH provided figures, tables (Figure 1A, Table 1), and contributed code for downstream analysis in the Freecyto software. IC co-wrote the manuscript and contributed to design of these studies. All authors read and approved the final manuscript.

\subsection*{Corresponding Author}
Correspondence to Nathan Wong (nathanwong@berkeley.edu) and Irina Conboy (iconboy@berkeley.edu). 

\section*{Ethics Declarations}
The authors declare no competing interests.

\section*{Additional Information}
\subsection*{Necessary Resources }
Freecyto is designed to be fully compatible with a standard user setup, and very little setup is required to begin using Freecyto for your flow cytometry needs. 
\begin{itemize}
    \item A web browser with JavaScript enabled (Core functions in the interactive analysis portion require JavaScript to be fully functional). Common browsers that satisfy this requirement include Google Chrome and Firefox. Mobile devices that have a mobile web browsing application can also satisfy this requirement.
    \item A valid Google ID or email address. This allows Freecyto to recognize the user and keep records of previous jobs performed under this user ID.
    \item A valid internet connection (HTTP, HTTPS) is required to access the online interface of Freecyto.
\end{itemize}
\subsection*{Walkthrough}
To begin, navigate to freecyto.com. Note that several documentation options are available for viewing on the home page. These options include: (1) Detailed, feature-specific documentation, (2) Video run-through of the application, (3) Open-source licenses and attributions (4) Freecyto’s privacy policy, and (5) Login URL to access the Freecyto application interface.
[Supplemental Figure 1]

\begin{flushleft}
Next, press “advanced analysis” to access the interactive visualizations of the flow cytometry data. This is an example of the shallow gating feature, in which selecting a sub-population of cells will display that sub-population across all selected fluorescence channels.
[Supplemental Figure 2]
\end{flushleft}

\bibliography{sample}
\newpage
\subsection*{Supplemental Figures}
\begin{figure}[h!]
\centering
\includegraphics[width=0.8\linewidth]{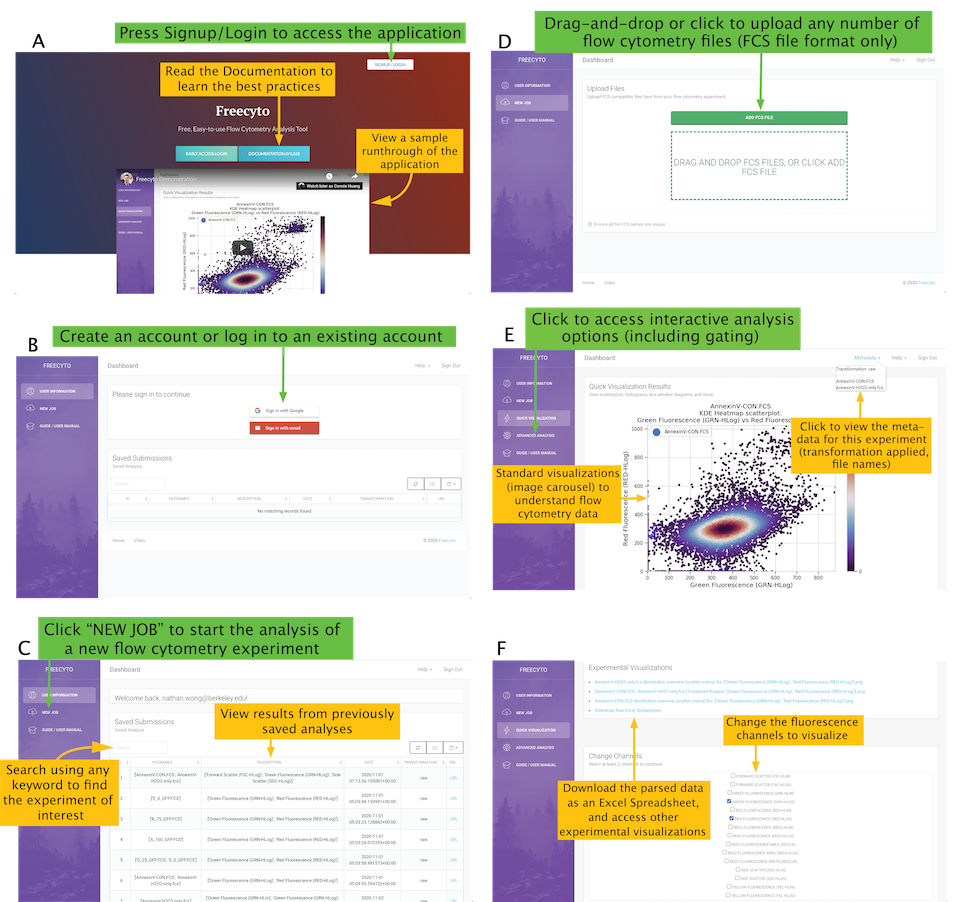}
\captionsetup{labelformat=empty}
\caption{\textbf{Supplemental Figure 1. Freecyto Quick Visualization Walkthrough.} \\\textbf{(A) Freecyto Homepage.} Navigate to freecyto.com and select login to continue. After clicking Login to access the Freecyto application interface, you need to create a new user account either through Google or email. If you already have an account on Freecyto, log in with those credentials. \\\textbf{(B) Freecyto Login Page.} Create an account using a Google or Email ID. Once you have successfully logged in, you will be able to access your personal user portal. From here, you can see all past analyses that you performed (linked to your user ID). You can also sort and search past saved analyses and access visualizations of those analyses directly and quickly by clicking on the corresponding link. \\\textbf{(C) Freecyto User Portal.} View previously performed analyses and access the page to create a new job. New users will have no previous experiments saved. However, each time the user uploads data or another user shares an experiment, the experiment will be listed in the table of the home page. These experiments can be sorted, indexed, and accessed without needing to repeat previously performed analysis operations. To begin a new job, click “New Job” located in the left column of the dashboard. Next, upload any number of FCS files you wish to analyze. \\\textbf{(D) Freecyto New Job.} Upload new FCS file(s) to begin a new analysis job. After the files have been uploaded, you will be able to access the quick visualizations page, in which the standard histograms, scatterplots, heatmaps, and box-whisker diagrams are displayed in a slideshow (image carousel) format. \\\textbf{(E) Freecyto Quick Visualization.} View histograms, scatterplots, box-whisker diagrams, heatmaps of the uploaded flow cytometry data. You may also change the fluorescence channels displayed at this time, by scrolling to the bottom of the page and selecting the new fluorescence channels to display. \\\textbf{(F) Changing the quick visualization display options.}}
\label{fig:Sf1}
\end{figure}

\newpage
\begin{figure}[h!t]
\centering
\includegraphics[width=1\linewidth]{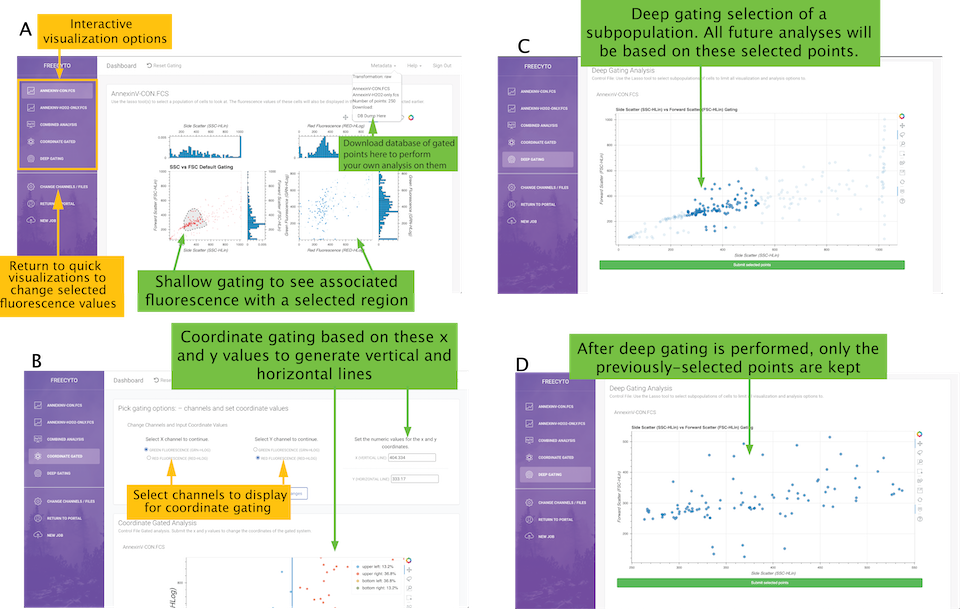}
\captionsetup{labelformat=empty}
\caption{\textbf{Supplemental Figure 2. Freecyto Interactive Analysis Walkthrough.} Next, press “advanced analysis” to access the interactive visualizations of the flow cytometry data. This is an example of the shallow gating feature, in which selecting a sub-population of cells will display that sub-population across all selected fluorescence channels. \\\textbf{(A) Freecyto Interactive Shallow Gating.} Shallow gating to see associated fluorescence values of a selected region. Coordinate gating analysis can then be performed to determine the percentage of cells that are located within or outside the bounds of preset x and y values.  \\\textbf{(B) Freecyto Interactive Coordinate Gating Display.} Gate flow cytometry experimental files based on specific X and Y values and see the percentage of cells within and outside these regions. Deep gating can also be performed to specifically examine sub-populations of cells.  \\\textbf{(C) Freecyto Interactive Deep Gating Display (Before).} \\\textbf{(D) Freecyto Interactive Deep Gating Display (After).}}
\label{fig:Sf2}
\end{figure}
\newpage

\begin{figure}[!ht]
\centering
\includegraphics[width=1\linewidth]{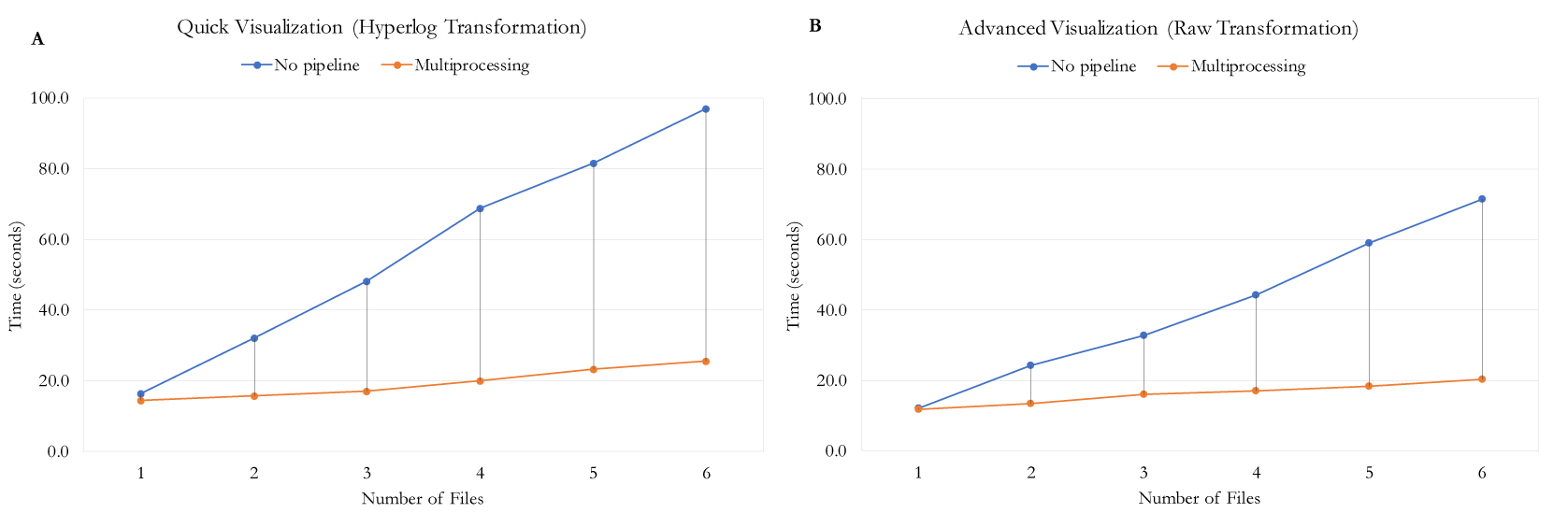}
\captionsetup{labelformat=empty}
\caption{\textbf{Supplemental Figure 3. Multiprocessing vs No Pipeline.} Plots show the time taken to process files when using multiprocessing vs. no multiprocessing for \textbf{(A)} Quick visualization and for \textbf{(B)} Advanced visualization.}
\label{fig:Sf3}
\end{figure}
\newpage
\begin{figure}[!ht]
\centering
\includegraphics[width=.8\linewidth]{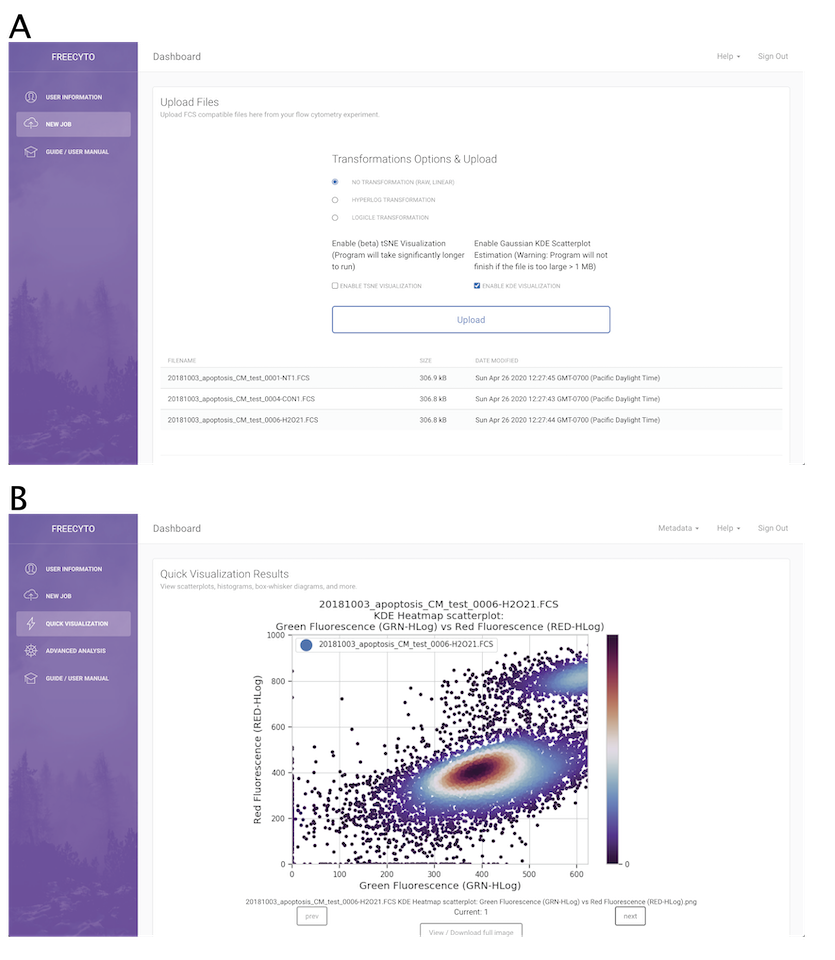}
\captionsetup{labelformat=empty}
\caption{\textbf{Supplemental Figure 4.} Some advantages of the Freecyto analysis include multiple file upload and quick data visualization. \\\textbf{(A) Multiple File Upload.} You can upload multiple files here and customize available settings, such as t-SNE and KDE visualizations with the option of various transformations. \\\textbf{(B) Quick Visualizations.} You now have access to many different visualizations of your uploaded data, including histograms, kernel density plots, and heatmaps.}
\label{fig:Sf4}
\end{figure}
\newpage
\begin{figure}[h!t]
\centering
\includegraphics[width=1\linewidth]{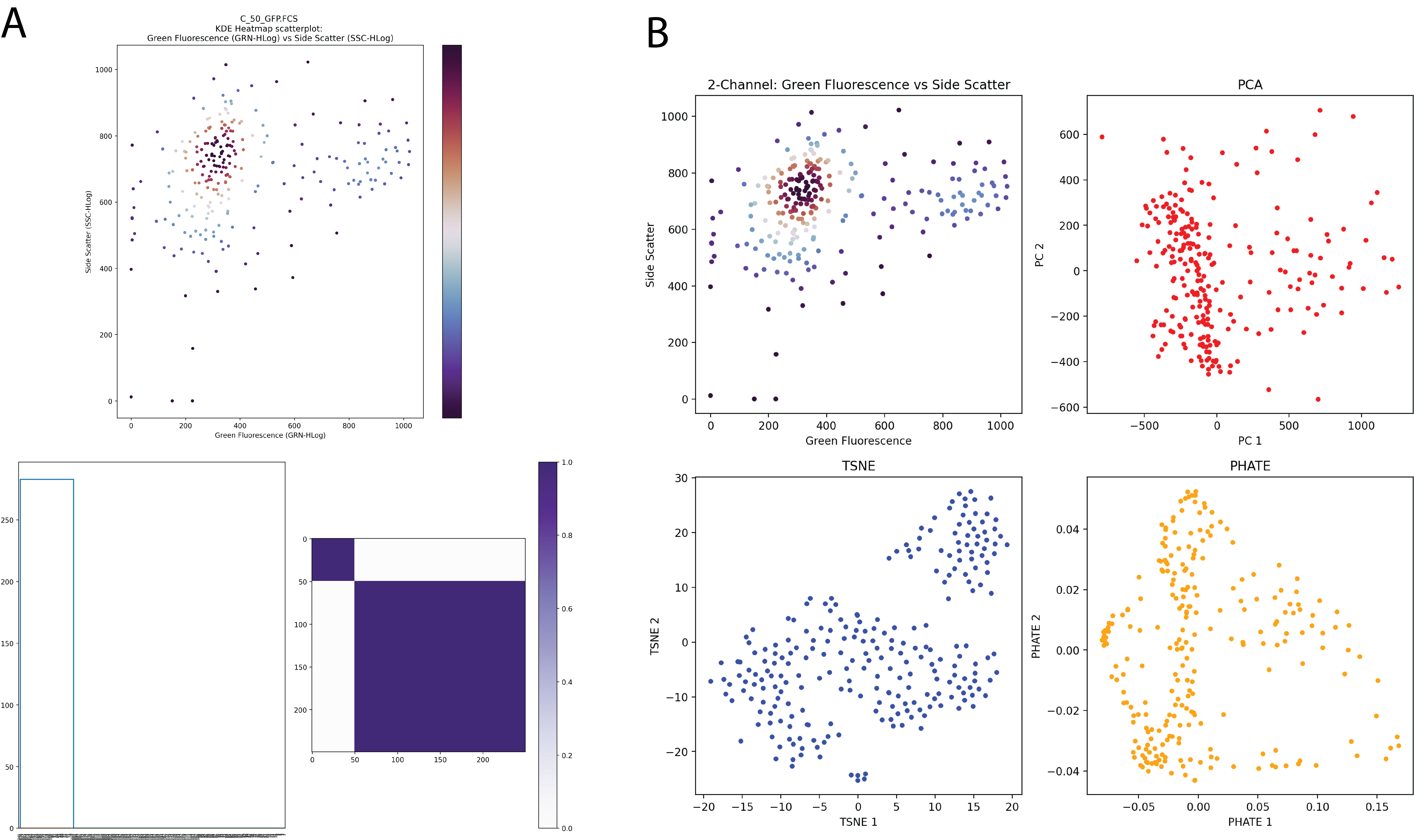}
\captionsetup{labelformat=empty}
\caption{\textbf{Supplemental Figure 5.} Downstream analysis of flow cytometry experiments. \\\textbf{(A) Visualizing the local structure of the 50:50 WT/GFP+ experiment.} Ward hierarchical clustering is performed downstream of the k-means quantization on the spearman correlation matrix of the Green Fluorescence and Side Scatter channels. We find the 2 distinct sub-populations as expected from this experiment. \\\textbf{(B) Dimensionality reduction comparison.} Various dimensionality reduction techniques (PCA, tSNE, PHATE\cite{Moon2019}) were performed on the same downstream data, but with all 15 channels selected as features. As expected, 2 distinct sub-populations were noted in each of these methods.}
\label{fig:Sf5}
\end{figure}
\end{document}